\documentclass[pdflatex,sn-mathphys-num]{sn-jnl-new}
\usepackage{graphicx}%
\usepackage{booktabs}%
\usepackage{siunitx}                       
\usepackage{amsmath,amssymb,amsfonts}
\usepackage{mathrsfs}

\DeclareSIUnit\px{px}

\usepackage{ragged2e}  
\usepackage[normalem]{ulem}
\usepackage{lineno}
\usepackage{caption}
\usepackage{subfigure}

\begin{document}

\title[]{Twist, turn and encounter: the trajectories of small atmospheric particles unravelled}

\author[1,2]{\fnm{T.} \sur{Bhowmick}}

\author[1]{\fnm{Y.} \sur{Wang}}\email{yong.wang@ds.mpg.de}

\author[3]{\fnm{J.} \sur{Latt}}

\author[1]{\fnm{G.} \sur{Bagheri}}\email{gholamhossein.bagheri@ds.mpg.de}

\affil[1]{
\orgname{Max Planck Institute for Dynamics and Self-Organisation}, 
\orgaddress{\street{Am Fa{\ss}berg 17},
\city{G\"ottingen},
\postcode{D-37077},
\country{Germany}}}

\affil[2]{
\orgname{University of G\"ottingen},
\orgaddress{\street{Friedrich-Hund-Platz 1},
\city{G\"ottingen},
\postcode{D-37077},
\country{Germany}}}

\affil[3]{
\orgname{University of Geneva},
\orgaddress{\street{24 rue du Général-Dufour},
\city{Genève},
\postcode{CH-1211},
\country{Switzerland}}}

\abstract{
Every solid particle in the atmosphere, from ice crystals and pollen to dust, ash, and microplastics, is non-spherical.
These particles play significant roles in Earth's climate system, influencing temperature, weather patterns, natural ecosystems, human health, and pollution levels.
However, our understanding of these particles is largely based on the theories for extremely small particles and experiments conducted in liquid mediums. 
In this study, we used an innovative experimental setup and particle-resolved numerical simulations to investigate the behaviour of sub-millimetre ellipsoids of varying shapes in the air.
Our results revealed complex decaying oscillation patterns involving numerous twists and turns in these particles, starkly contrasting their dynamics in liquid mediums.
We found that the frequency and decay rate of these oscillations have a strong dependence on the particle shape.
Interestingly, disk-shaped particles oscillated at nearly twice the frequency of rod-shaped particles, though their oscillations also decayed more rapidly.
During oscillation, even subtly non-spherical particles can drift laterally up to ten times their volume-equivalent spherical diameter.
This behaviour enables particles to sweep through four times more air both vertically and laterally compared to a volume-equivalent sphere, significantly increasing their encounter rate and aggregation possibility.
Our findings provide an explanation for the long-range transport and naturally occurring aggregate formation of highly non-spherical particles such as snowflakes and volcanic ash.
}


\maketitle

\begin{figure}%
\centering
\includegraphics[width=\textwidth]{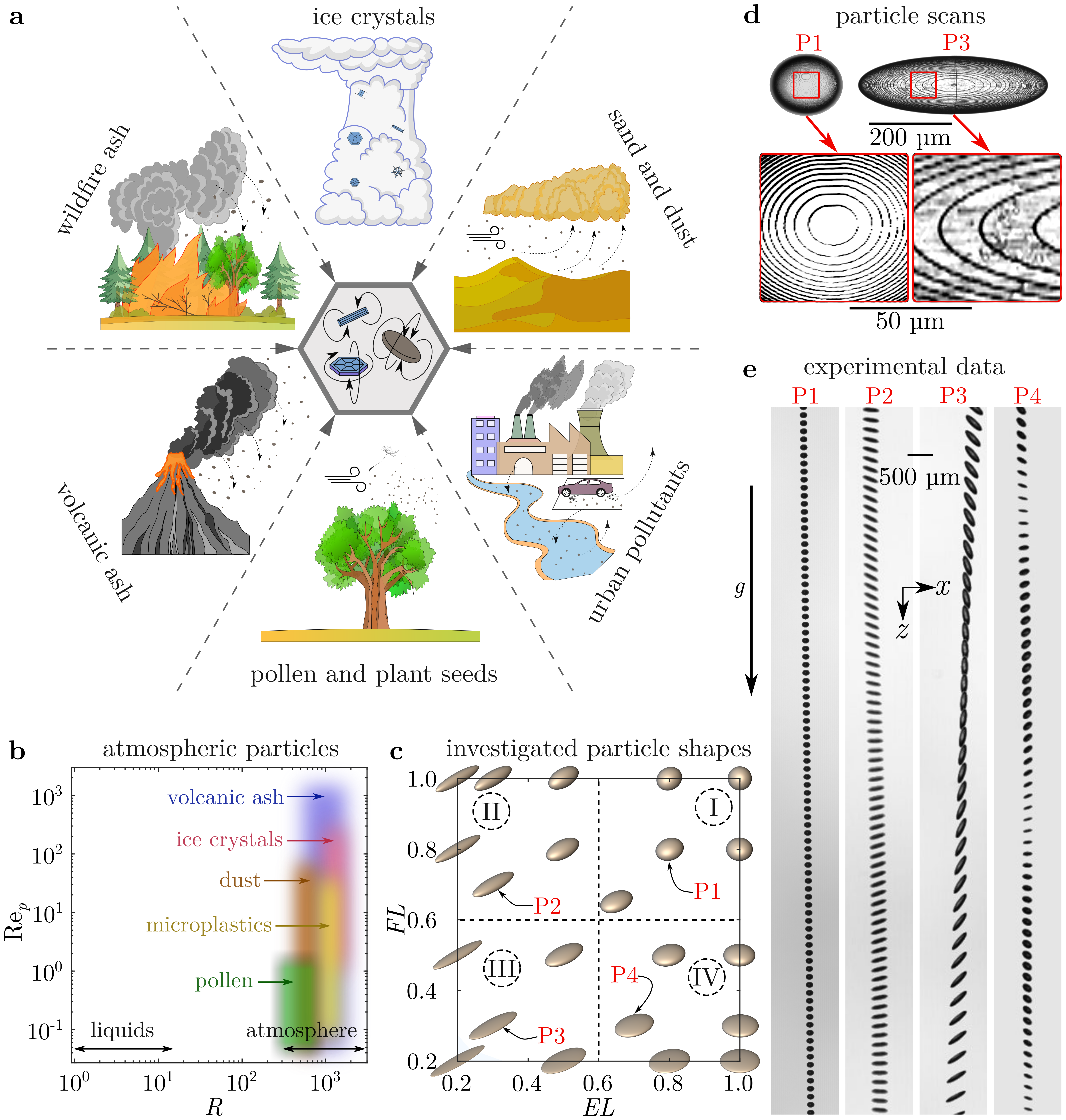}
\caption{
\textbf{Atmospheric non-spherical particles and their behaviour.}
\textbf{a}, Natural and anthropogenic sources of non-spherical solid particles that impact the weather, climate and ecosystems.
\textbf{b}, Atmospheric particles in the phase space of particle Reynolds number ${\rm Re}_p$ and particle to fluid density ratio $R$ \cite{Kajikawa:1972,Kajikawa_1992,Pruppacher_Klett_2010,Bagheri_Bonadonna_2016,vanHout:2004,Allen:2019,Zhang:2020}.
\textbf{c}, The shapes of ellipsoidal, non-spherical particles of this study are defined with two shape parameters -- elongation $EL$ and flatness $FL$, keeping the volume constant (equivalent to a sphere of $\SI{140}{\micro\meter}$ diameter $D_{eq}$).
The particle shape-space is divided in four categories: I -- near spherical, II -- prolate / elongated, III -- highly anisotropic ellipsoidal, and IV -- oblate / flattened shapes.
\textbf{d}, Laser microscope scans of two 3D printed particles from categories I and III.
The 3D surface features are $<\SI{1}{\micro\meter}$ in size. 
\textbf{e}, Experimental observation of oscillations in angular orientation and lateral drift when non-spherical particles falling under gravity $g$.
}
\label{fig:Fig1}
\end{figure}

There are numerous examples of non-spherical atmospheric particles from both natural and anthropogenic sources.
Examples include volcanic \cite{Rossi_2021} and wildfire ashes \cite{Bodi_2014}, sand \cite{Merrison_2012}, dust \cite{Mahowald_2014,van_der_Does_2018} and microplastics \cite{Brahney_2020,Tatsii_2023} (Fig. \ref{fig:Fig1}a).
Those particles are known to have various impacts, from increasing health risks for humans \cite{Horwell_2006,Manisalidis_2020,Zhang_2020} to the radiative properties of clouds \cite{Baran_2012,Matus_2017}, which influence the weather and climate \cite{Ramanathan_1989,Norris_2016}.
However, our understanding of the transport properties of these particles is rather limited.
This is due to the lack of theoretical models for heavy particles larger than 0.1 mm in air and the existing experimental data come almost exclusively from studies conducted in liquids.
The experiments in liquids \cite{willmarth1964steady,Jayaweera_Cottis_1969,field1997chaotic,Lopez_Guazzelli_2017,Roy_2019,Esteban_2020,Cabrera_2022,Baker_Coletti_2022,Roy23,Giurgiu_2024}, in which the dynamic similarity is observed using the particle Reynolds number ${\rm Re}_p$, provide many insightful findings on the settling behaviour of non-spherical particles.
Here ${\rm Re}_p$ quantifies the relative importance of the inertial forces compared to the viscous forces acting on the particle \cite{Happel_Brenner_1983,Khayat_Cox_1989,Gustavsson_2019}.
However, the ratio of particle to fluid density, $R = \rho_p /\rho_f$, which in air is typically of the order of 1000 for most solid particles, cannot exceed much beyond 10 in liquids.
Recent advancements have enabled experimental studies of low Reynolds numbers in air, particularly for axisymmetric shapes like spheroids.
A combination of analytical tools and empirical corrections explains that the particle inertia significantly impacts their settling dynamics and has the potential to trigger strong oscillations in the transient behaviour of such particles  \cite{bhowmick_PRL_2024}.
The settling motion and rotational dynamics of non-axisymmetric particles, however, remain poorly understood.
An instinctive question here is: given the significant difference in density between non-spherical atmospheric particles and air, but their typically low Reynolds number, how do these particles settle and what are the consequences?

This study focuses on the range $1 < {\rm Re}_p < 10$ at $R = 1000$, which is exactly the parameter space for many sources of atmospheric particles (Fig. \ref{fig:Fig1}b). 
The absence of controlled experiments in this critical parameter space stands as a testimony to the significant experimental and numerical challenges involved.
Here we investigate the intriguing dynamics of heavy ellipsoids of various shapes as they fall in still air.
Ellipsoids have been demonstrated as highly effective surrogates for irregular particles \cite{Bagheri_Bonadonna_2016}, and this allows us to systematically examine the dynamics of particles across a wide range of shapes.
Our work integrates novel experimental measurements with particle-resolved direct numerical simulations (DNS) to reveal the complex dynamics of heavy solid particles falling in air. 
This study illuminates that the transient dynamics of small non-spherical particles in air represent a largely unexplored field and provides significant implications for the understanding of our particle-rich atmospheric environment.

\begin{figure}[htb!]%
\centering
\includegraphics[width=1.0\textwidth]{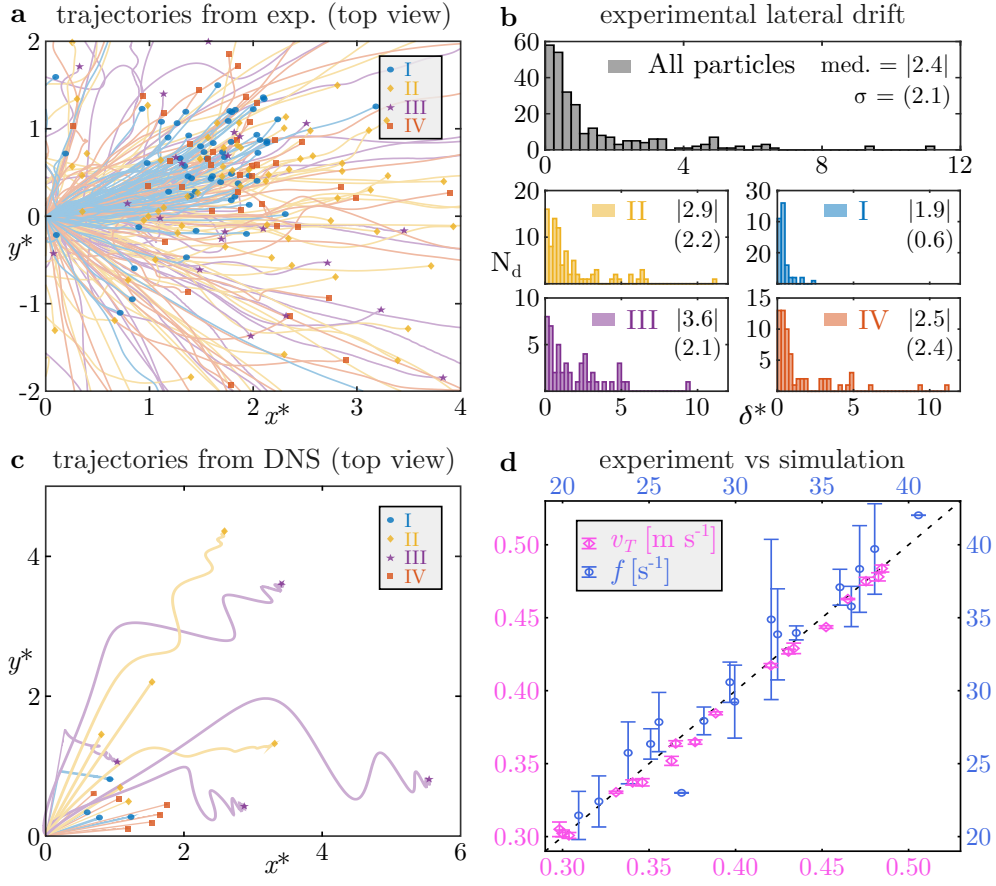}
\caption{
\textbf{Drifting oscillating settling pattern.}
\textbf{a}, Top view of the drifting trajectories of the non-spherical particles from the experiments (zoomed in for better visibility).
$x^*$ and $y^*$ are the two lateral components normalised by $D_{eq}$.
Particle categories I-IV are from Fig. \ref{fig:Fig1}c.
\textbf{b}, Histograms of the normalised lateral drift ($\delta^* = \delta/D_{eq}$) with a bin width of $0.25\,\delta^*$.
$\delta$ is calculated as the integral of lateral displacements between two consecutive camera frames.
$N_d$ is the number of experimental runs for a specific amount of $\delta^*$, and the median and standard deviation values are shown as med. and $\sigma$.
\textbf{c}, Top view of the drifting particle trajectories from the DNS (rotated for better visibility), where the particles are released with zero initial velocity and minimum projection area normal to the falling direction.
\textbf{d}, Comparison of the terminal velocities $v_T$ and frequencies of the angular oscillations $f$ of the particles from experiments (vertical axes) and DNS (horizontal axes).
The error bars are equivalent to one standard deviation variation.
The dashed diagonal line shows 1:1 correspondence between experiments and DNS.
The relative differences between the experiments and DNS are less than $3\%$ and $10\%$, respectively, for the $v_T$ and $f$.
}
\label{fig:Fig2}
\end{figure}

The particles of this study are of ellipsoidal shape, defined by four shape categories (I-IV) and two shape parameters: Elongation ($EL = I/L$) and Flatness ($FL = S/I$) (Fig. \ref{fig:Fig1}c), where $L$, $I$, and $S$, are longest, intermediate and shortest dimensions of the particle and orthogonal to each other (Extended Data Table 1).
The control parameter in our experiment is the particle mass (or volume), which is equal to that of a volume-equivalent sphere with a diameter $D_{eq} = \SI{140}{\micro\meter}$ for all particles. 
The particles are produced using two-photon polymerisation 3D printing, which enables us to achieve an explicit control over particle shape and size with sub-micrometer printing resolution (Fig. \ref{fig:Fig1}d).
Particle density $\rho_p$ is \SI{1200}{\kg\per\meter\cubed} \cite{Liu_2018}. More experimental details can be found in Methods.
We have conducted a total of 262 successful experimental runs for different particle shapes, with each shape having from 9 to 22 runs.
By covering a wide range of flattened and/or elongated ellipsoids, we comprehensively explore the phase space of diverse particle shapes.
This approach also serves as an effective proxy for irregular particles with similar form \cite{Bagheri_Bonadonna_2016}, eliminating the need for infinite case-by-case investigations.
The particle Reynolds number achieved here ranges from approximately ${\rm Re}_p =2.1$ to $4.5$, where ${\rm Re}_p = D_{eq} v_T/\nu$, with $v_T$ as the terminal velocity and $\nu$ as the kinematic viscosity of air. 
This is much lower than the threshold Reynolds number of about 100, beyond which an unstable falling pattern is observed due to hydrodynamic instabilities in the wake of the falling particles in air and in liquids \cite{willmarth1964steady,Ern_2012,Auguste_2013,Esteban_2020,Tinklenberg_2023}.
A coin tossed in water or a falling leaf are typical examples of non-spherical particles with ${\rm Re}_p>100$.

Our experiments show that ellipsoids of various shapes exhibit orientation oscillations and lateral drifts as they settle in still air (Fig. \ref{fig:Fig1}e).
The more flattened (low $FL$) or elongated (low $EL$) the ellipsoids are, the more pronounced these effects become.
The top view of the particle trajectories (Fig. \ref{fig:Fig2}a) and the histograms of lateral drift (Fig. \ref{fig:Fig2}b) reveal the intricate falling behaviour of the investigated particles.
Across all shape categories, particularly in categories II-IV, there is a notable lateral drift exceeding $10 D_{eq}$.
Some particles exhibit wavy lateral trajectories, while some others also make multiple sharp turns (Fig. \ref{fig:Fig2}a).
The elongated particles in categories III and II have the highest median lateral drift, while the standard deviation of lateral drift in categories II-IV is four times higher than the near spherical particles in category I (Fig. \ref{fig:Fig2}b).

To fully understand the dynamics behind the complex trajectory of these particles, we have replicated the experiments with large-scale DNS (see Methods).
All simulations are carried out with a fixed initial condition, i.e. release with zero initial velocity and minimum projection area normal to the falling direction.
Details of the novel DNS can be found in \cite{Bhowmick_2024_arxiv2}. 
Top view of DNS trajectories also reveal the wavy lateral motion of the particles (Fig. \ref{fig:Fig2}c), with sharp turns and lateral drifts up to $6 D_{eq}$ for category III, compared to only one $D_{eq}$ lateral drift for category I.
The excellent quantitative agreement between the experiments and DNS for the terminal velocity $v_T$ and oscillation frequency $f$ of the particles (Fig. \ref{fig:Fig2}d) suggests that many features of the transient dynamics are independent of initial conditions and could be generalised to real-world scenarios.
For each shape, the standard deviation of $v_T$ is less than $2\%$ of its magnitude and in case of $f$ is less than $16\%$, despite a large variability expected in the initial conditions of the experimental trajectories.

\begin{figure}[htb!]%
\centering
\includegraphics[width=\textwidth]{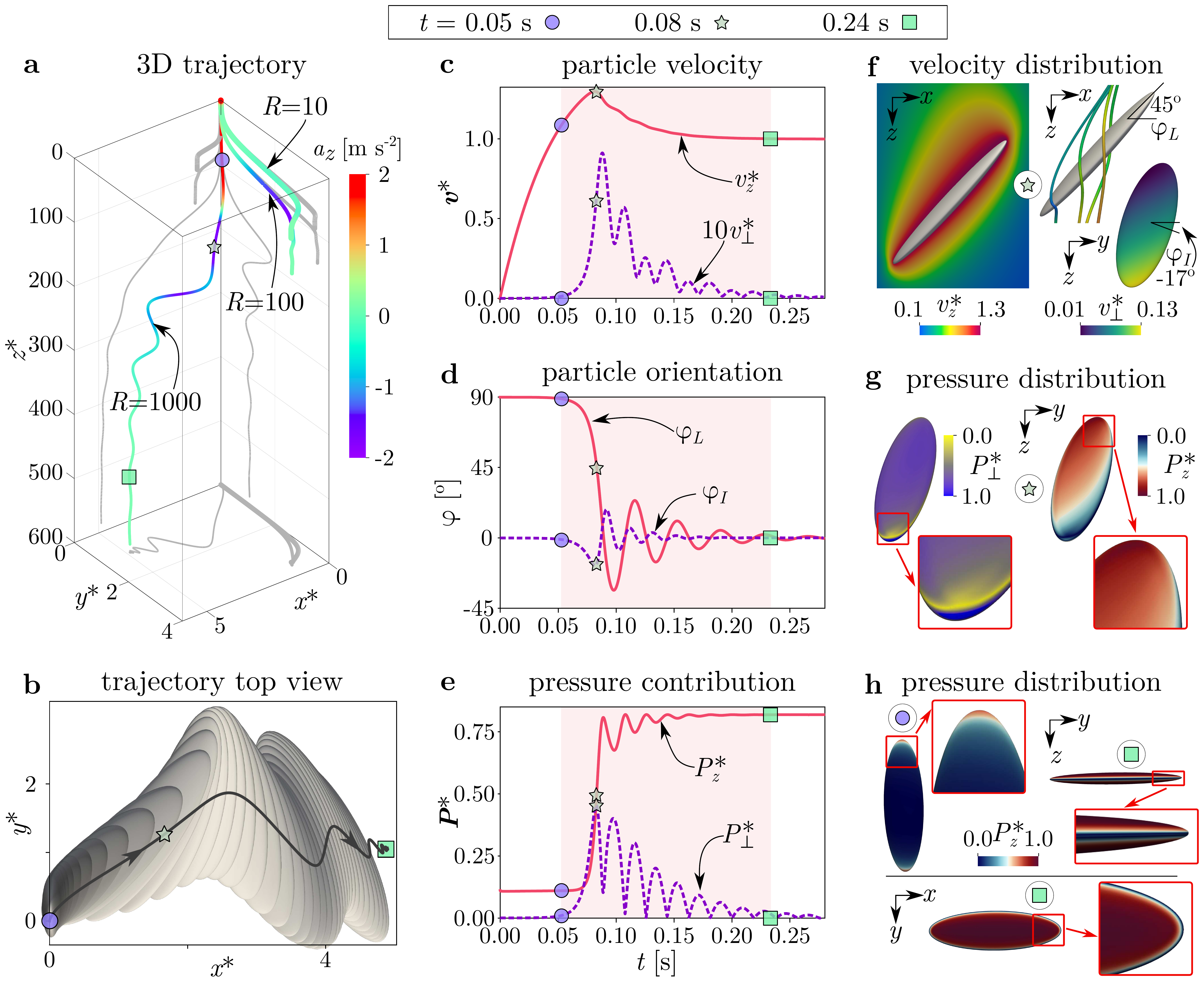}
\caption{
\textbf{Settling dynamics of \emph{EL}=\emph{FL}=0.3 particles in air and liquids.}
\textbf{a}, Three dimensional particle trajectories (coloured by the instantaneous acceleration $a_z$) and their two dimensional projections (grey) at three different density ratios $R$ = 1000, 100, 10, but same particle Reynolds number ${\rm Re}_p=3$.
Different transient phases are marked by: circle -- start of angular motion, star -- a pitch angle $\varphi_L$ ($L$-axis angle from horizon) of \SI{45}{\degree}, and square -- onset of terminal state.
\textbf{b}, Top view of the $R$ = 1000 particle trajectory together with the superimposed particle images.
Time $t$ evolution of the $R$ = 1000, \textbf{c}, particle velocity normalised by terminal velocity $\boldsymbol{v}^* = \boldsymbol{v}/v_T$
($v_z^*$ is the vertical component, and $v_\perp^*$ is the magnitude of the lateral components),
\textbf{d}, particle orientation ($\varphi_L$ is the pitch angle and $\varphi_I$ is the roll angle -- $I$-axis angle from horizon),
\textbf{e}, particle's pressure contribution $\boldsymbol{P}^* = \boldsymbol{P}/F_z$ normalised by the the vertical component of the drag force $F_z$
($P_z^*$ and $P_\perp^*$ are respectively the vertical component and the magnitude of the lateral components).
At $\varphi_L=\SI{45}{\degree}$ ($\varphi_I=\SI{-17}{\degree}$), \textbf{f}, the velocity distribution and streamlines of airflow around the particle (coloured based on the $v_z^*$), and the surface distribution of the $v_\perp^*$; and \textbf{g}, the surface distribution of the $P_\perp^*$ and $P_z^*$.
\textbf{h}, The surface distribution of the $P_z^*$ at $\varphi_L=\SI{0}{\degree}$ and \SI{90}{\degree}.
}
\label{fig:Fig3}
\end{figure}

Having successfully validated our simulations against experimental measurements, we turn our attention to the particle with $EL=FL=0.3$.
This particle serves as a representative case for all other shapes, allowing us to thoroughly detail its full dynamic behaviour.
We firstly compare the three-dimensional trajectories of this ellipsoid in liquids of three different densities, but with the same particle Reynolds number, ${\rm Re}_p = 3$ (Fig. \ref{fig:Fig3}a).
The density ratios $R$ considered here are: 1000 for most solid particles in air, and 100 and 10 for, e.g.,  particles made of heavy metal falling in SF\textsubscript{6} (pressurised to 15 bars) and in water, respectively.
It is observed that particles reach the terminal state at larger vertical distances, $z^* = z/D_{eq}$, with increasing $R$.
For example, the required $z^*$ to reach the terminal state at $R = 1000$ is about 6.3 (3.6) times longer than that at $R = 10$ ($R = 100$).
In addition, the $R=1000$ particle trajectory exhibits twists and turns (see the superimposed particle images from top view in Fig. \ref{fig:Fig3}b), in stark contrast to the particle trajectories at other density ratios.
The transition to the terminal state as well as the oscillations in the particle orientation and the forces acting on the particle display marked deviations between the $R = 1000$ case and other cases (see Extended Data Fig. 1).
These discrepancies suggests the existence of an entirely unique dynamical behaviour for non-spherical atmospheric particles.

The transient motion of the particle with $R=1000$ proceeds in three different phases, whereby the initial accelerating phase is followed by an angular oscillation phase, which is damped over time and finally reaches the terminal state (Fig. \ref{fig:Fig3}c-e).
During the initial phase (until the purple circle at \SI{0.05}{\second} in Fig. \ref{fig:Fig3}), when the particle retains its initial orientation while accelerating, neither lateral drift nor lateral velocity develops (Fig. \ref{fig:Fig3}c).
Both pitch and roll angles, $\varphi_L$  ($L$-axis angle from horizon) and $\varphi_I$  ($I$-axis angle from horizon), retain their initial values of \SI{90}{\degree} and \SI{0}{\degree}, respectively (Fig. \ref{fig:Fig3}d).
Up to this point, both the vertical and lateral components of the pressure contributions to the forces exerted on the particle (Fig. \ref{fig:Fig3}e) are at their minimum, i.e. most of the force exerted by the fluid on the particle is caused by skin friction.(Fig. \ref{fig:Fig3}h).

The onset of angular oscillation phase begins beyond the purple circle at \SI{0.05}{s}, coinciding with rapid changes in both roll and pitch angles (Fig. \ref{fig:Fig3}d), as well as a sharp increase in lateral velocity (up to 14\% of the vertical terminal velocity, Fig. \ref{fig:Fig3}c).
The proportion of pressure forces in the total force exerted on the particle also increases (Fig. \ref{fig:Fig3}e).
The particle's vertical velocity continues to rise and overshoots up to 30\% beyond the terminal velocity as the particle surpasses the pitch angle of $\varphi_L = \SI{45}{\degree}$ (Fig. \ref{fig:Fig3}c).
At this orientation (the grey star in Fig. \ref{fig:Fig3}), the local lateral velocity on the particle surface shows a marked uneven distribution with significant variability, while no separation or re-circulation zone is observed in the particle wake (Fig. \ref{fig:Fig3}f).
Meanwhile, the lateral pressure contribution reaches a peak magnitude of $\sim 50\%$ similar to the vertical pressure contribution (Fig. \ref{fig:Fig3}e).
Simultaneously, the local pressure contribution on the particle surface develops an asymmetrical distribution (Fig. \ref{fig:Fig3}g), resulting in torques in both lateral and vertical directions.
From there the particle begins to twist and turn noticeably, and approaches its terminal state in a non-linear manner, i.e. with oscillations in velocity, orientation and the forces acting on it.

The angular oscillations in particle orientation (Fig. \ref{fig:Fig3}d) have two different frequencies, a lower frequency of 26.9 Hz for the pitch angle $\varphi_L$ and a higher frequency of 51.8 Hz for the roll angle $\varphi_I$.
The orientation amplitudes attenuate with an exponential decay rate $\mu$, which is 21.2 Hz for pitch and 40.2 Hz for roll angles.
The oscillations in $v_\perp^*$ (Fig. \ref{fig:Fig3}c), as well as in $P_z^*$ and $P_\perp^*$ (Fig. \ref{fig:Fig3}e) occur with the same frequency as of the pitch angle.
The roll angle, on the other hand, has no discernible influence on the oscillations of the $v_\perp^*$, $P_z^*$ and $P_\perp^*$.
In addition, while $P_\perp^*$ oscillates in phase with the $\varphi_L$, $v_\perp^*$ and $P_z^*$ oscillate $\SI{90}{\degree}$ out of phase with the $\varphi_L$, i.e., $P_\perp^*$ increases when $P_z^*$ decreases.
At terminal state (after the green square at 0.24~s in Fig. \ref{fig:Fig3}), the particle stops angular oscillation with largest projection area perpendicular to the direction of gravity.
Thus the parameters keep constant, i.e. $v_z^* = 1.0$ and $v_\perp^* = 0.0$ (Fig. \ref{fig:Fig3}c), $\varphi_L = \varphi_I = \SI{0}{\degree}$ (Fig. \ref{fig:Fig3}d), and $P_z^* = 0.82$ and $P_\perp^* = 0.0$ (Fig. \ref{fig:Fig3}e).
The fluid-particle dynamics at different density ratios exhibit significantly different transients (Extended Data Fig. 1).
This further emphasises that complex trajectories and recurring angular oscillations at low particle Reynolds numbers (at an absence of separation or re-circulation zone at ${\rm Re}_p\approx1$) occur only at high density ratios.

\begin{figure}[tb]%
\centering
\includegraphics[width=\textwidth]{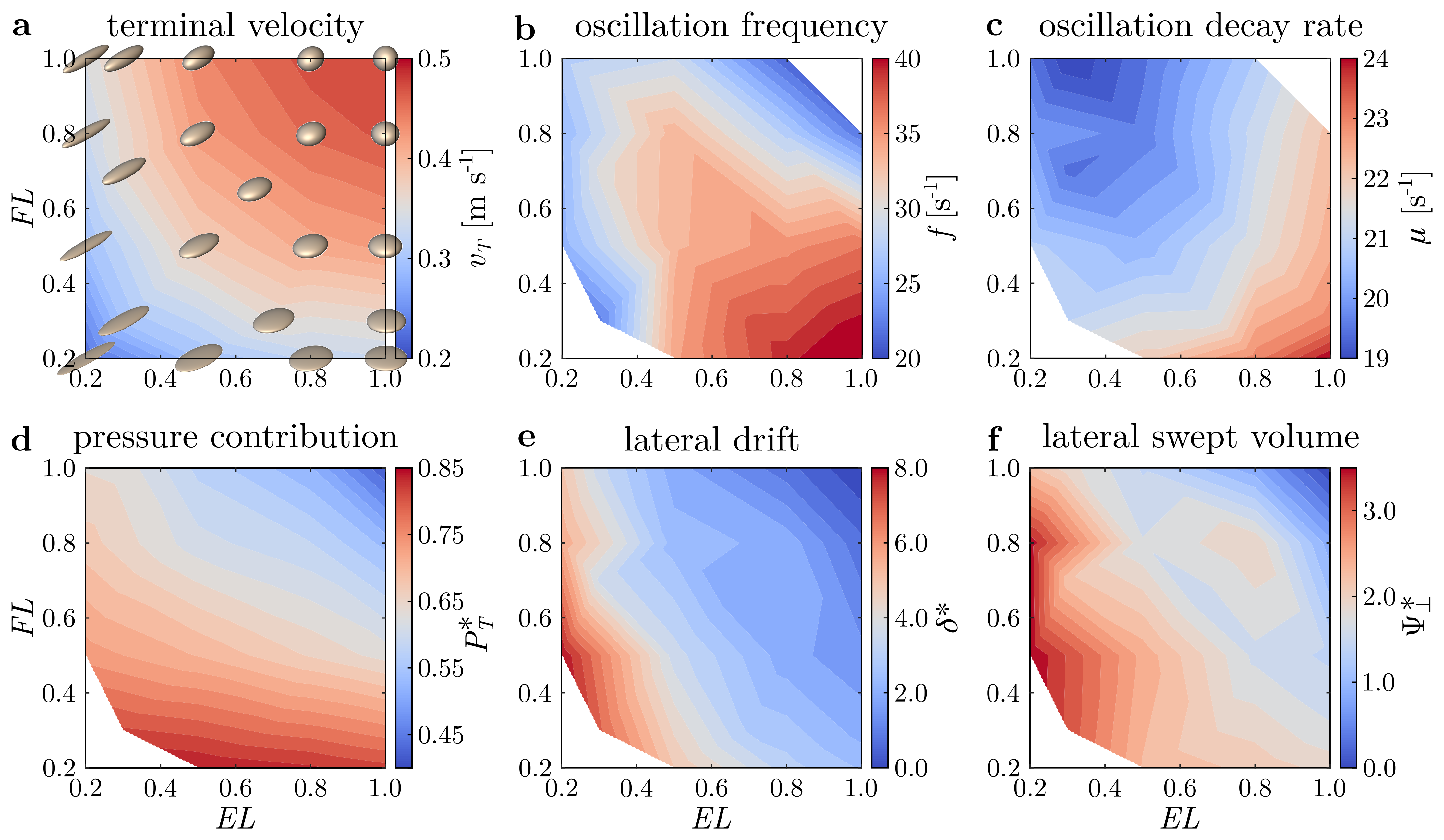}
\caption{
\textbf{Phase diagrams of important quantities.} 
In the phase space of particle elongation $EL$ and flatness $FL$, the distributions of \textbf{a}, the terminal velocity $v_T$, \textbf{b}, the oscillation frequency of the angular orientation $f$, \textbf{c}, the decay rate of the oscillation amplitudes $\mu$, \textbf{d}, the terminal state pressure contribution in the vertical direction $P_T^*$, \textbf{e}, the normalised lateral drift $\delta^*$, and \textbf{f}, the normalised lateral swept volume $\Psi_\perp^*$ are shown.
The scattered data are linearly interpolated using griddata in Matlab (R2023b).
In the transient motion of a sphere, $f$ and $\mu$ are not present, which are shown as white triangles in the upper right corner of \textbf{b} and \textbf{c}.
For ellipsoid $EL=FL=0.2$, only $v_T$ was measured.
The diagrams in \textbf{a} and \textbf{b} show values from the experiments, while \textbf{c-f} are from the DNS.
Detailed tabulated data are in Extended Data Tables 2-5.
}
\label{fig:Fig4}
\end{figure}

Using simulations and experiments, we also systematically investigate important particle quantities as a function of the particle shape parameters $EL$ and $FL$ (Fig. \ref{fig:Fig4}).
Tabulated data of such quantities is provided in the Extended Data Tables 2-5. 
The terminal velocity $v_T$ shows a $2.5$ fold change in its magnitude (Fig. \ref{fig:Fig4}a).
The sensitivity of the terminal velocity to elongation and flatness is very similar, resembling  that of ellipsoids in the Stokes regime \cite{Bagheri_Bonadonna_2016}.
The oscillation frequency $f$ of the angular orientation of the particles on the other hand, is strongest for flattened and non-elongated particles with up to 2 times more oscillations (Fig. \ref{fig:Fig4}b).
We have already showed that the amplitudes of the angular oscillations decays over time in an exponential manner (Fig. \ref{fig:Fig3}c).
The exponential decay rate of the oscillation amplitudes $\mu$ shows that it is also higher for flattened and non-elongated particles (up to 1.3 times) but in a different manner than that of $f$(Fig. \ref{fig:Fig4}c).
The $v_T$, $f$ and $\mu$ are observed to be independent of the initial conditions of the particles, and therefore the results of the DNS agree closely with the experiments (Extended Data Tables 2-4).

The other quantities, such as lateral drift and swept volume, provide a fair comparison when performed under a fixed initial condition, which is achieved by DNS.
In addition, the distribution of forces and secondary oscillations around the intermediate axis can only be analysed using DNS results. 
The terminal state contribution of the vertical component of pressure force to that of the vertical drag force, $P_T^*$, exhibits strong dependency on flatness (Fig. \ref{fig:Fig4}d).
While 43\% of the total drag force exerted on the spherical particle ($EL=FL=1.0$) is caused by the pressure forces, for an ellipsoid with $EL=FL=0.3$ it is almost twice as much at around 82\%.
In contrast, the integral of the lateral displacements -- the lateral drift, which is normalised by particle diameter $\delta^* = \delta/D_{eq}$, exhibits strong dependency on elongation (Fig. \ref{fig:Fig4}e).
This agrees with the experimental results (Fig. \ref{fig:Fig2}b), although the initial condition of the particles cannot be precisely controlled.
A wide range of lateral drift is observed in DNS, with no drift for the sphere ($EL=FL=1.0$) and up to $8D_{eq}$ for the particle $EL=0.2$ and $FL=0.5$.
The volume that a particle disperses laterally to reach its terminal state is defined as the lateral swept volume.
The lateral swept volume normalised by the particle volume, $\Psi_\perp^*$, is highest for the elongated ellipsoidal particles.  It can be up to $3.5$, while it is 0 for a sphere (Fig. \ref{fig:Fig4}f).
The normalised vertical swept volume $\Psi_z^*$ depends largely on the largest projection area of the particle, i.e. on the orientation in the terminal state, and increases from 1 for a sphere to 3.7 for the particles considered here (Extended Data Table 5).

Our results show that even for moderately elongated and/or flattened non-spherical particles, settling dynamics in air are dramatically different from those in liquids despite having identical Reynolds numbers. 
Spherical particles exhibit no lateral drift, whereas the non-spherical particles considered here can drift by a factor of 10 times their volume-equivalent spherical diameter in air.
This behaviour is confirmed by both experimental measurements with varying initial conditions and numerical simulations with a fixed initial condition.
Furthermore, the increase in swept volume, which directly correlates with particle collision or encounter rate \cite{Saffman_Turner_1956,Meyer_2011,Wilczek2022PNAS}, accelerates aggregation formation as the particles become increasingly non-spherical.
In nature, the enhancements in lateral transport, particle encounters, aggregation and the resulting effects on the atmospheric residence time are likely to be even greater than shown in this study, given the abundance of particles with significantly non-spherical shape in the atmosphere \cite{Magono_1966,Um_2015,Bagheri_Bonadonna_2016,Zhang:2020,Xiao_2023,Tatsii_2023}.
For example, typical elongation and flatness of natural ice crystals varies between 0.01 to 0.5 \cite{Auer_1970}, while microplastics have very low values of flatness and elongation, with a length of \SI{1}{\micro\meter} to \SI{5}{\milli\meter} and a thickness of \SI{1}{\micro\meter} to \SI{0.5}{\milli\meter} \cite{Zhang:2020,Tatsii_2023}.
Such shape results in significantly higher swept volumes and lateral drift values than those reported here.
The strong lateral drift combined with increased swept volume also explains the long-range transport and the formation of loose aggregate of volcanic ashes \cite{BAGHERI2016520,Brown_2012}, snowflakes \cite{Leinonen_2020,Grazioli_2022,Maahn_2024}, and microplastics \cite{Dissanayake_2022} in nature.
Turbulence additionally affects the orientation of a non-spherical particle \cite{pumir2016collisional,Voth_Soldati_2017} and amplifies the settling instabilities \cite{bhowmick_PRL_2024}, making the settling dynamics of atmospheric particles more complex and interlinked.

\section*{Acknowledgements}
T. B. was funded by the German Research Foundation (DFG) Walter Benjamin Position (Project No. 463393443). This work was supported by the Max Planck Society. Scientific activities are carried out in Max Planck Institute for Dynamics and Self-Organization (MPIDS). Computational resources from the Max Planck Computing and Data Facility and MPIDS are gratefully acknowledged.

\section*{Author contributions}
Conceptualization and research design: T.B., Y.W. and G.B.; experiments: T.B.; postprocessing of the experimental data: T.B. and G.B.; simulations: T.B.; postprocessing of the simulation data: T.B. and Y.W.; analysis of the results: T.B., Y.W. and G.B.; development of simulation software: T.B., Y.W. and J.L.; visualization: T.B., Y.W. and G.B.; interpretation of results: T.B., Y.W. and G.B.; writing-original draft: T.B., Y.W., J.L. and G.B; writing-review and editing: T.B., Y.W., J.L. and G.B.

\bibliography{refs_v2}

\end{document}